\documentclass{sf2a-conf2012}
\usepackage{graphicx}
\usepackage{hyperref}
\usepackage[]{natbib}  
\usepackage[cyr]{aeguill}
\usepackage{epstopdf}

\def\BibTeX{{\rm B\kern-.05em{\sc i\kern-.025em b}\kern-.08em
    T\kern-.1667em\lower.7ex\hbox{E}\kern-.125emX}}
\bibpunct{(}{)}{;}{a}{}{,}  

\begin{document}

\TitreGlobal{SF2A 2012}


\title{A new method based on Markov chains for deriving SB2 orbits directly from their spectra}

\runningtitle{A BiT MOrE with MCMC}

\author{J.-B. Salomon}\address{Observatoire Astronomique de Strasbourg, UMR 7550, 11, rue de l'universit\'e, F-67000 Strasbourg, France}

\author{R. Ibata$^{1}$}

\author{P. Guillout$^{1}$}

\author{J.-L. Halbwachs$^{1}$}

\author{F. Arenou}\address{GEPI, Observatoire de Paris-Meudon, F-92195 Meudon Cedex, France}

\author{B. Famaey$^{1}$}

\author{Y. Lebreton$^{2}$}

\author{T. Mazeh}\address{School of Physics and Astronomy, Raymond and Beverly Sackler Faculty of Exact Sciences, Tel Aviv University, Tel Aviv, Israel}

\author{D. Pourbaix}\address{FNRS, Universit\'e libre de Bruxelles, CP226, boulevard du Triomphe, 1050 Bruxelles, Belgium}

\author{L. Tal-Or$^{3}$}




\setcounter{page}{237}


\maketitle


\begin{abstract}
We present a new method to derive orbital elements of double-lined spectroscopic binaries (SB2). The aim is to have accurate orbital parameters of a selection of SB2 in order to prepare the exploitation of astrometric Gaia observations. Combined with our results, they should allow one to measure the mass of each star with a precision of better than 1 \%.

The new method presented here consists of using the spectra at all epochs simultaneously to derive the orbital elements without templates. It is based on a Markov chain including a new method for disentangling the spectra.
\end{abstract}



\section{Introduction}

Double-lined spectroscopic binaries (SB2) form the basis of the least model--dependent methods used to derive stellar masses. Their orbital elements are employed to derive the products ${\cal M}_* \sin^3 i$, where ${\cal M}_*$ is the mass of a component and $i$ is the inclination of the orbital plane. Therefore, when the inclination may be obtained from another technique, such as eclipse observations or astrometric measurements, the accuracy of the masses depends on that of the SB2 orbital elements.
 
The derivation of the elements of a spectroscopic orbit is usually based on the derivation of the radial velocities (RV) of the components from the blended spectra of the binary star. This is generally done by fitting one \citep{Hill93} or two spectral templates to each spectrum \citep{Zucker94}. The approach is justifiable when the templates are similar to the actual spectra of the components, but otherwise the results are degraded by template mis-match.
 
Other methods do not use templates, but are based on the disentangling of the spectra \citep{SimStu94, GonzLev06}. The method presented hereafter falls into this category. It is Bayesian, since it is assumed that the radial velocities obey Keplerian motions, and the orbital elements are derived at the same time that the spectra are disentangled. It is based on Markov chains and on an new disentangling technique. The method is described in Sect.~\ref{method} hereafter. In Sect.~\ref{hd}, it is applied to HD~89745.

\section{A BiT MOrE with MCMC}\label{method}

A BiT MOrE with MCMC is an \textbf{A}lgorithm for \textbf{Bi}naries \textbf{T}o \textbf{M}atch \textbf{Or}bital \textbf{E}lements with \textbf{M}arkov \textbf{C}hain \textbf{M}onte \textbf{C}arlo. It combines a MCMC routine and a disentangling method.

\subsection{The MCMC method}

Our method is based on a Markov chain Monte Carlo (MCMC) survey of model parameter space (see, e.g., \citealt{MacKay03}), which samples the free parameters and produces a posterior probability density function (PDF) for each dimension. To get that, the MCMC routine investigates randomly the space of parameters. Thus, it is well adapted for our case with 6 orbital elements (the systemic velocity is lost, see next section).

The MCMC algorithm requires an initial guess for the elements of the spectroscopic orbit: the amplitudes of the RV of the two components, the orbital period, the eccentricity, the epoch and the argument of the periastron. Subsequently, the MCMC routine solves the orbital equations to derive the RV for each time of observation.

\subsection{The disentangling method}

The goal of this step is to build a model from the orbital elements. For that, we developed a new spectral disentangling algorithm acting in Fourier space in order to extract each spectral component. It is different from iterative procedures like \cite{GonzLev06} or \citet{1995A&AS..114..393H} because it does not use references line or least-square fitting.

Actually, the spectra ($S$) are the sum of the two components ($M_A$ and $M_B$) which are shifted ($\delta x_A$ and $\delta x_B$) according to their orbital RV at the time of observation $j$ (equation~\ref{spectra}). 
\begin{equation}
S = M_A(x-\delta x_A^j) + M_B(x-\delta x_B^j) \label{spectra}
\end{equation}
The Doppler shift is used to disentangle the two components in Fourier space where the gap between each epoch is just a phase shift. The Fourier transform is calculated using Fast Fourier Transform techniques (e.g., \citealt{1992nrfa.book.....P}). So, it is possible to determine, right from the initialisation step, $M_A$ and $M_B$. Then, the two disentangled spectra are summed to obtain the model spectra. Note that information on the values of continua are lost in passing to Fourier space. Moreover, the systemic velocity is not derived because only RV shifts are involved.

\subsection{Likelihood}

Finally, the model is compared to the actual spectrum to test the likelihood of the solution. It is injected again at the beginning of the algorithm to find the next parameters. This procedure is reiterated until the Markov chain converges.

\subsection{Validation}

To validate our method, synthetic spectra have been built assuming a set of orbital elements. In order to be sure that the method is reliable, the computation was started assuming initial elements quite different from those used in the simulation. The algorithm was run to attempt to recover the input parameter values and their uncertainties.

\section{Application}\label{hd}

Our method was applied to a set of 11 spectra of HD 89745, taken over 1785 days, corresponding to about 77~\% of the period. These spectra arise out of a large observation program at the Haute-Provence observatory with the T193/Sophie \citep{Halbwachs09}, and also from another program carried out with the same instrument \citep{Halbwachs12}. This star was chosen for the number of spectra, although it appears somewhat atypical. In the past, it received 27 Coravel observations, and an orbit was derived after adding 7 Sophie spectra (Halbwachs et al. 2012). For this calculation, all the RV were derived from the cross-correlation function (CCF) of the spectra with a single mask or a simple template, the 1-mask technique. However, when the TODMOR algorithm was applied to a subset of 9 Sophie spectra, the two-template CCF yielded RVs quite different from those obtained with a single mask. The consequence was that the RV semi-amplitudes, $K_1$ and $K_2$, were quite different from the previous values: around 11 and 8~km/s instead of 7 and 11~km/s, respectively. The mass ratio of the binary is then changed from 0.62 to 1.30. If this discrepancy is due to a bias in the RV estimations obtained through a single template, one would expect to find it again with a disentangling technique.

\begin{figure}[ht!]
 \centering
 \includegraphics[width=0.8\textwidth,clip]{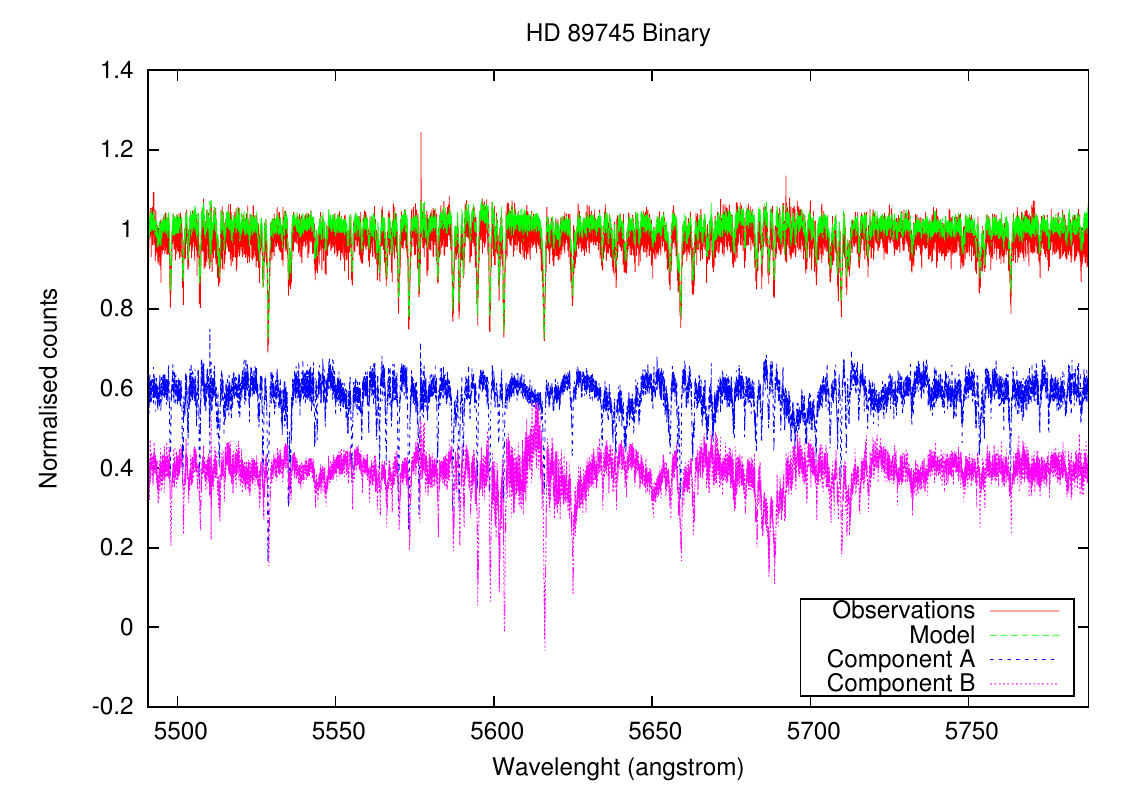}      
  \caption{Observation of HD 89745 in red, component A in blue and B in pink, model built (sum of the two components) in green.}
  \label{salomon:fig1}
\end{figure}

\begin{figure}[ht!]
 \centering
 \includegraphics[width=0.8\textwidth,clip]{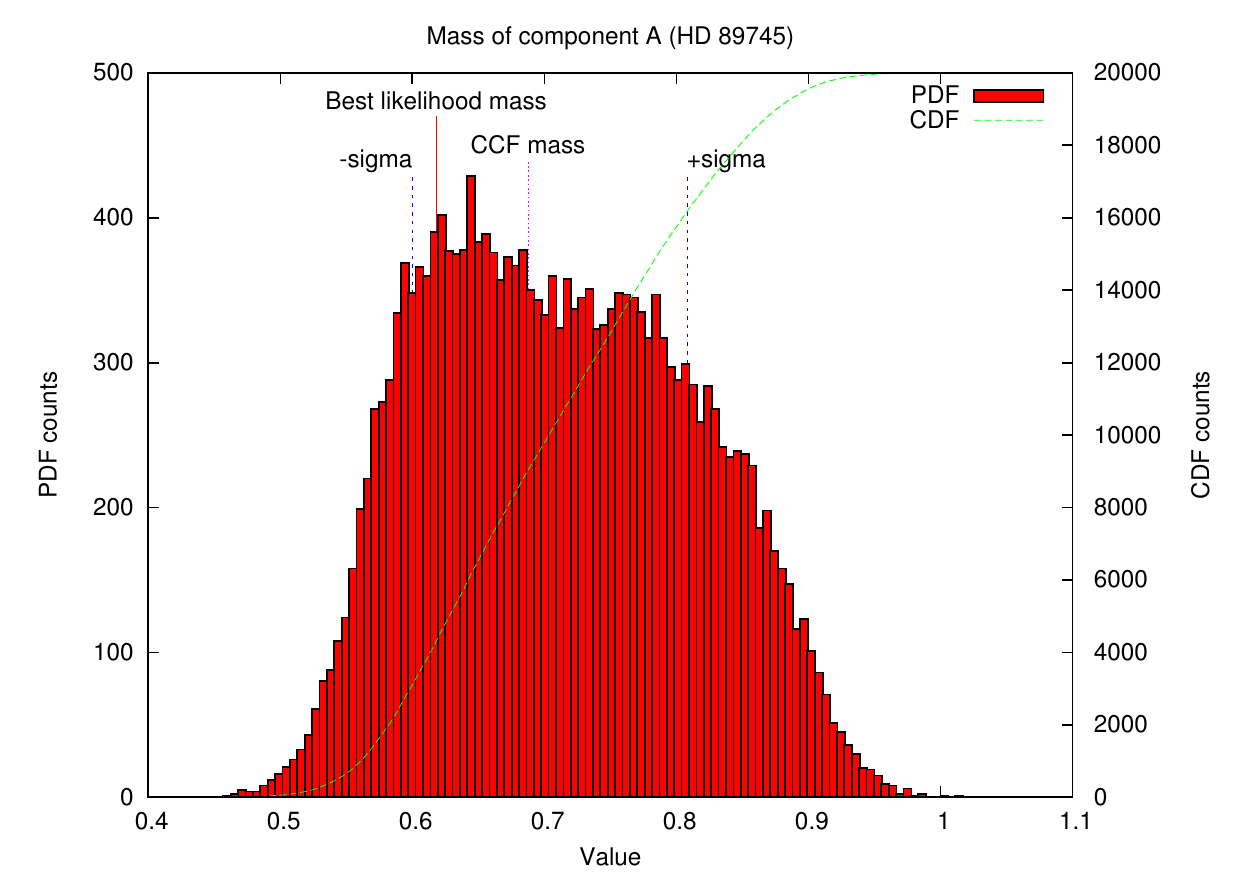}      
  \caption{Probability density function in red and cumulative density function in green of HD 89745 component A mass (${\cal M}_1 \times sin^3i$). The best likelihood value is indicated in red with its uncertainty in blue and the mass obtained by cross-correlation \citep{Halbwachs12} in pink.}
  \label{salomon:fig2}
\end{figure}

A small wavelength range, just 3 orders, is taken into account between 5490.8 and 5787.9 Angstr\"{o}ms to avoid telluric lines (for a the present assessment). All spectra have a good signal to noise ratio (higher than 50).
The disentangling of the two components and their associated model are shown for one observation in Fig.~\ref{salomon:fig1}. It was obtained after only 20000 iterations in A BiT MOrE with MCMC. From the PDF (see Fig.~\ref{salomon:fig2}), the uncertainties at one sigma around the best likelihood value can be obtained (listed in table~\ref{tab:mesure}).

\begin{table}
\centering
\caption{Obital elements of HD 89745 obtained by double peak cross-correlation, 1-mask technique \citep{Halbwachs12}, column 1, by TODMOR, column 2 and by A BiT MOrE with MCMC, column 3. The elements cannot be compared since the input observations are different for each of the three techniques, see explanations in the text.}
\vspace{0.5 cm}
\begin{tabular}{lccc}
  \hline\hline
    Orbital elements & \cite{Halbwachs12} & TODMOR                        & A BiT MOrE with MCMC \\
 \hline
    $K_1$ (km/s)    & $6.983\pm{0.183}$   &  $10.735\pm{0.137}$           & $8.700_{-1.641}^{+0.380}$   \\
    $K_2$ (km/s)    & $11.184\pm{0.197}$  &  $8.252\pm{0.146}$            & $9.420_{-0.250}^{+1.177}$   \\
    e               & $0.3902\pm{0.0176}$ &  $0.1197\pm{0.0120}$          & $0.432_{-0.099}^{+0.004}$   \\
    P  (jours)      & $2303.6\pm{10.5}$   &  $2555.684570\pm{152.240479}$ & $2437.8_{-87.1}^{+68.9}$ \\
    $T_0$ (jours)   & $52093\pm{26.8}$    &  $54405.2383\pm{176.9055}$    & $52230.25_{-239.426}^{+1.256}$ \\
    $\omega$  (deg) & $174.8\pm{4.8}$     &  $175.03\pm{28.87}$           & $142.5_{-1.7}^{+12.3}$  \\
 \hline\hline
    Other elements\\
 \hline
    $M_1sin^3{i}$ ($M_{\odot}$)  & $0.6876\pm{0.030}$ & $0.772649\pm{0.053345}$ & $0.618_{-0.018}^{+0.190}$ \\
    $M_2sin^3{i}$ ($M_{\odot}$)  & $0.4293\pm{0.022}$ & $1.005062\pm{0.068305}$ & $0.640_{-0.234}^{+0.014}$ \\
 \hline

\end{tabular}
\label{tab:mesure}
\end{table}

The values obtained with A BiT MOrE with MCMC are roughly in agreement with the previous values. We have large uncertainties due to the small number of spectra, the incomplete period coverage and the fact that only three orders are taken into account. However, we confirm that the mass ratio could be inversed, as found with the TODMOR reduction. It is worth noticing that this does not come from a permutation between the primary component and the secondary one, since the orbital elements $T_0$ and $\omega$ are in agreement with those found with the 1-mask technique. Nevertheless, these results are only preliminary and more observations are needed to draw reliable conclusions.

\section{Conclusion}

The simulations have shown that A BiT MOrE with MCMC is a powerful new tool to perform orbital element calculations. This method should provide reliable orbital elements and recover the spectra of the individual stars at the same time.

We have also seen that the orbital elements obtained from a single template are only roughly consistent with those obtained when two templates are used or when the spectra are disentangled. For HD 89745, using two templates and disentangling both leads to orbital elements rather different from those obtained from a single template, and we plan to derive a new orbit when more spectra become available.

However, more detailed investigations are required to compare the reliability of the three techniques considered here. This will be done in the future in order to obtain the most accurate masses when Gaia measurements are available.

\begin{acknowledgements}
Thanks to AS Gaia for functioning support. The observations were successful thanks to the help of the staff of the OHP. It is a pleasure to thank Eric Gosset for valuable discussions.
\end{acknowledgements}

\bibliographystyle{aa}  
\bibliography{salomon} 

\end{document}